\documentclass[conference]{IEEEtran}

\usepackage{amssymb}

\newcommand{\A}{{\mathcal{A}}}

\newcommand{\CC}{{\mathcal{C}}}

\newcommand{\E}{{\mathcal{E}}}
\newcommand{\FF}{{\mathcal{F}}}

\newcommand{\HH}{{\mathcal{H}}}

\newcommand{\SSS}{{\mathcal{S}}}
\newcommand{\T}{{\mathcal{T}}}

\newcommand{\C}{{\mathbb{C}}}
\newcommand{\F}{{\mathbb{F}}}

\newcommand{\Z}{{\mathbb{Z}}}
\newcommand{\zerob}{{\mathbf 0}}
\newcommand{\ab}{{\mathbf a}}

\newcommand{\fb}{{\mathbf f}}

\newcommand{\tb}{{\mathbf t}}

\newcommand{\wb}{{\mathbf w}}

\newcommand{\Wb}{{\mathbf W}}

\newcommand{\ie}{{\em i.e., }}
\newcommand{\eg}{{\em e.g., }}
\newcommand{\cf}{\emph{cf.\ }}
\newcommand{\etal}{\emph{et al.\ }}

\newcommand{\Tr}{\mathrm{Tr~}}

\newcommand{\half}{\frac{1}{2}}

\newcommand{\dfree}{d_{\mathrm{free}}}

\newcommand{\openbox}{\leavevmode
     \hbox to.77778em{%
     \hfil\vrule
     \vbox to.675em{\hrule width.6em\vfil\hrule}%
     \vrule\hfil}}
\newcommand{\qed}{\hspace*{1cm}\hspace*{\fill}\openbox}

\begin{document}

\title{MacWilliams Identities for \\ Terminated Convolutional Codes}

\author{G. David Forney, Jr. \\
Laboratory for Information and Decision Systems\\
Massachusetts Institute of Technology\\
Cambridge, MA 02139\\
Email: forneyd@comcast.net}

\maketitle

\begin{abstract}
Shearer and McEliece \cite{SM77} showed that there is no MacWilliams identity for the free distance spectra of orthogonal linear convolutional codes.  We show that on the other hand there does exist a MacWilliams identity between the generating functions of the weight distributions per unit time of a linear convolutional code $\CC$ and its orthogonal code $\CC^\perp$, and that this distribution is as useful as the free distance spectrum for estimating code performance.  These observations are similar to those made recently by Bocharova \etal \cite{BHJK};  however, we focus on terminating by tail-biting rather than by truncation.
\end{abstract}

\section{Introduction}

Finding a MacWilliams-type identity for convolutional codes is a problem of long standing \cite{SM77}.
For a linear time-invariant convolutional code $\CC$ over a finite field, the most commonly studied distance distribution is the \emph{free (Hamming) distance spectrum}, namely, the distribution of (Hamming) weights of codewords in $\CC$ that start and end in the zero state without passing through an intermediate zero state.  Shearer and McEliece \cite{SM77} showed by example that the free distance spectrum of $\CC$ does not in general determine that of $\CC^\perp$, and therefore that there could be no MacWilliams identity for such distributions.

Gluesing-Luerssen and Schneider (GLS) have recently formulated  \cite{GLS08} and proved  \cite{GLS09} a MacWilliams-type identity for convolutional codes involving the Hamming weight adjacency matrix (HWAM) of a convolutional code $\CC$ and the HWAM of its orthogonal code $\CC^\perp$.  In \cite{F09, F10}, the GLS result was proved in a different way, and generalized to various kinds of weight adjacency matrices and to group codes defined on graphs.

More recently, Bocharova, Hug, Johannesson and Kudryashov \cite{BHJK} have proved a different MacWilliams-type identity for truncations of a convolutional code $\CC$ and its orthogonal code $\CC^\perp$, and have shown that by letting the truncation length become large, an approximation to the free distance spectrum can be obtained.
In this paper, which is mostly based on \cite{F10}, we derive similar results for weight distributions of block codes obtained by various kinds of termination procedures, of which we regard tail-biting as the nicest.  We argue that these alternative distributions are just as useful for estimating code performance as the free distance spectrum.  These results effectively answer the original question posed by Shearer and McEliece \cite{SM77}.  


\section{Terminated convolutional codes}

A general method for approximating the free distance spectrum of a linear convolutional code $\CC$ is to derive a series of block codes $\CC_N$ of length $N$ from $\CC$ by some sort of termination procedure, and then to study the distance distributions of $\CC_N$ as $N \to \infty$.  As we shall see in Section III, for any of the termination methods below, the distance distribution of $\CC_N$, normalized by $N$, approaches the free distance spectrum of $\CC$ for $\dfree \le d < 2\dfree$.   However, the usual termination methods are problematic if we are also interested in the distance distribution of the orthogonal convolutional code $\CC^\perp$.  

For example, the most common termination method is to take the \emph{subcode} $\CC_{[0,N)}$ of $\CC$, consisting of all sequences in $\CC$ whose support is contained in the interval $[0, N) = \{k \in \Z \mid 0 \le k < N\}$ (\ie all code sequences that pass through the zero state at times 0 and $N$, restricted to $[0,N)$), which is effectively a block code of length $N$ time units.  As a subcode of $\CC$, $\CC_{[0,N)}$ has at least the minimum (free) distance of $\CC$.  However, the orthogonal code to $\CC_{[0,N)}$ is the \emph{projection} $(\CC^\perp)_{|[0,N)}$ of the orthogonal convolutional code $\CC^\perp$ onto the interval $[0, N)$  (\ie all orthogonal code sequences that pass through any state at times 0 and $N$, restricted to $[0,N)$).  In general, a projection $(\CC^\perp)_{|[0,N)}$  has low-weight codewords, no matter how large $N$ becomes.

Bocharova \etal \cite{BHJK} have considered another kind of terminated code that they call a \emph{truncated code}, which we will denote by $\CC_{\lhd [0,N)}$.  Such a code may be described as the subcode $\CC_{[0,\infty)}$ projected onto $[0,N)$:  \ie $$\CC_{\lhd [0,N)} = (\CC_{[0,\infty)})_{|[0,N)}.$$  By projection/subcode duality, the dual code $(\CC_{\lhd [0,N)})^\perp$  is
$$
((\CC_{[0,\infty)})_{|[0,N)})^\perp = ((\CC_{[0,\infty))})^\perp)_{[0,N)} = ((\CC^\perp)_{|[0,\infty))})_{[0,N)};
$$
\ie the subcode defined on $[0,N)$ of the projection of the orthogonal convolutional code $\CC^\perp$ onto $[0, \infty)$.  Alternatively,  $(\CC_{\lhd [0,N)})^\perp$ may be defined as a reverse-truncated code \cite{BHJK}  $$(\CC^\perp)_{\rhd [0,N)} = ((\CC^\perp)_{(-\infty,N))})_{|[0,N)}.$$  Thus there is a MacWilliams identity between the weight distributions of $\CC_{\lhd [0,N)}$ and $(\CC^\perp)_{\rhd [0,N)}$.  Moreover, unlike the subcode $\CC_{[0,N)}$ or the projection $\CC_{|[0,N)}$, a truncated code has the same rate as $\CC$.  However, truncated codes have  small minimum distance.

A more elegant method of terminating a convolutional code $\CC$ is via tail-biting.  The \emph{tail-biting terminated code} $\CC_{||[0,N)}$ of the convolutional code $\CC$ on the interval $[0, N)$ is the set of all codewords in $\CC$ that pass through the same state at times 0 and $N$, restricted to $[0,N)$.  For large enough $N$, the tail-biting code $\CC_{||[0,N)}$ has the same minimum distance as $\CC$;  moreover,  $\CC_{||[0,N)}$ has the same rate as $\CC$.  Most importantly, the orthogonal code to $\CC_{||[0,N)}$ is $(\CC^\perp)_{||[0,N)}$, the tail-biting terminated code of $\CC^\perp$ on the same interval \cite{F01}.  Thus there is a MacWilliams identity between the weight distributions of $\CC_{||[0,N)}$ and $(\CC^\perp)_{||[0,N)}$.  Finally, we will see that these distributions approach the free distance spectra of $\CC$ and $\CC^\perp$ nicely as $N \to \infty$.

\vspace{1ex}
\noindent
\textbf{Example 1} (rate-1/2 4-state binary linear convolutional code).  Consider the rate-1/2 binary linear time-invariant convolutional code $\CC$ with degree-2 generators $(1 + D^2, 1 + D + D^2)$, in standard $D$-transform notation.  A minimal encoder for this code is the linear time-invariant system with impulse response $(11, 01, 11, 00, \ldots)$, which has the 4-state trellis section shown in Figure 1(a).  The orthogonal convolutional code $\CC^\perp$ is the rate-1/2 binary linear convolutional code with  generators $(1 + D + D^2, 1 + D^2)$, which has a minimal 4-state linear encoder with impulse response $(11, 10, 11, 00, \ldots)$, and the 4-state trellis section shown in Figure 1(b).

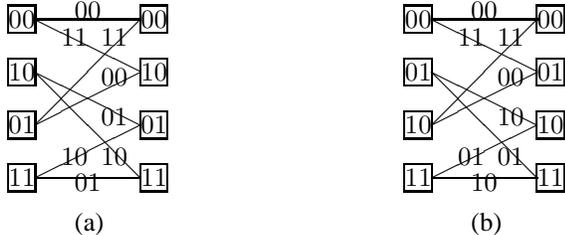
\begin{figure}[h]
\setlength{\unitlength}{5pt}
\centering
\begin{picture}(42,16)(0, -1)
\put(0,0){\framebox(2,2){$11$}}
\put(0,4){\framebox(2,2){$01$}}
\put(0,8){\framebox(2,2){$10$}}
\put(0,12){\framebox(2,2){$00$}}
\put(10,0){\framebox(2,2){$11$}}
\put(10,4){\framebox(2,2){$01$}}
\put(10,8){\framebox(2,2){$10$}}
\put(10,12){\framebox(2,2){$00$}}
\put(2,1){\line(1,0){8}}
\put(2,1){\line(2,1){8}}
\put(2,5){\line(2,1){8}}
\put(2,13){\line(1,0){8}}
\put(2,9){\line(1,-1){8}}
\put(2,13){\line(2,-1){8}}
\put(2,9){\line(2,-1){8}}
\put(2,5){\line(1,1){8}}
\put(5,0){$01$}
\put(4,2){$10$}
\put(7,5){$01$}
\put(7,2){$10$}
\put(5,13){$00$}
\put(4,11){$11$}
\put(7,8){$00$}
\put(7,11){$11$}
\put(5,-3){(a)}


\put(30,0){\framebox(2,2){$11$}}
\put(30,4){\framebox(2,2){$10$}}
\put(30,8){\framebox(2,2){$01$}}
\put(30,12){\framebox(2,2){$00$}}
\put(40,0){\framebox(2,2){$11$}}
\put(40,4){\framebox(2,2){$10$}}
\put(40,8){\framebox(2,2){$01$}}
\put(40,12){\framebox(2,2){$00$}}
\put(32,1){\line(1,0){8}}
\put(32,1){\line(2,1){8}}
\put(32,5){\line(2,1){8}}
\put(32,13){\line(1,0){8}}
\put(32,9){\line(1,-1){8}}
\put(32,13){\line(2,-1){8}}
\put(32,9){\line(2,-1){8}}
\put(32,5){\line(1,1){8}}
\put(35,0){$10$}
\put(34,2){$01$}
\put(37,5){$10$}
\put(37,2){$01$}
\put(35,13){$00$}
\put(34,11){$11$}
\put(37,8){$00$}
\put(37,11){$11$}
\put(35,-3){(b)}

\end{picture}

\caption{Trellis sections of (a) rate-1/2 4-state binary convolutional code $\CC$; (b) orthogonal code $\CC^\perp$.}
\label{Fig3}
\end{figure}

We now consider various methods of terminating this convolutional code $\CC$ with a block length of $N = 4$.  The subcode $\CC_{[0,4)}$ is the $(8,2)$ binary linear block code generated by the two generators
$$
\begin{array}{cccc}
11 & 01 & 11 & 00 \\
00 & 11 & 01 & 11 \\
\end{array}
$$
The minimum distance of this block code is the same as that of $\CC$, namely 5, although its rate is lower.
The orthogonal code to the subcode $\CC_{[0,4)}$ is the projection $(\CC^\perp)_{|[0,4)}$ of the orthogonal convolutional code $\CC^\perp$, which is the  $(8,6)$ binary linear block code generated by the six generators
$$
\begin{array}{cccc}
11 & 00 & 00 & 00 \\
10 & 11 & 00 & 00 \\
11 & 10 & 11 & 00 \\
00 & 11 & 10 & 11 \\
00 & 00 & 11 & 10 \\
00 & 00 & 00 & 11 \\
\end{array}
$$
The minimum distance of this block code is 2, less than that of $\CC^\perp$, although its rate is higher.

The truncated code $\CC_{\lhd[0,4)}$ is the $(8,4)$ binary linear block code generated by
$$
\begin{array}{cccc}
11 & 01 & 11 & 00 \\
00 & 11 & 01 & 11 \\
00 & 00 & 11 & 01 \\
00 & 00 & 00 & 11 \\
\end{array}
$$
The minimum distance of this block code is 2, but its rate is the same as that of $\CC$.  Its orthogonal code  $(\CC^\perp)_{\rhd[0,4)}$ is the $(8,4)$ binary linear block code generated by
$$
\begin{array}{cccc}
11 & 00 & 00 & 00 \\
10 & 11 & 00 & 00 \\
11 & 10 & 11 & 00 \\
00 & 11 & 10 & 11 \\
\end{array}
$$
which has the same parameters.

The tail-biting terminated code $\CC_{||[0,4)}$ is the $(8,4)$ binary linear block code generated by
$$
\begin{array}{cccc}
11 & 01 & 11 & 00 \\
00 & 11 & 01 & 11 \\
11 & 00 & 11 & 01 \\
01 & 11 & 00 & 11  
\end{array}
$$
whereas the orthogonal tail-biting terminated code $(\CC^\perp)_{||[0,N)}$ is the $(8,4)$ binary linear block code generated by the four generators
$$
\begin{array}{cccc}
11 & 10 & 11 & 00 \\
00 & 11 & 10 & 11 \\
11 & 00 & 11 & 10 \\
10 & 11 & 00 & 11  
\end{array}
$$
Both of these codes have a minimum distance of only 2 (\eg for paths such as 01 00 01 00 from state 10 to state 10).  However, for $N \ge 10$, it turns out that the minimum distance of both tail-biting terminated codes is 5, the same as the minimum distance of $\CC$ or $\CC^\perp$. \qed \vspace{1ex}

\section{Free distance spectra for convolutional codes from terminated codes}

Let us now consider how the free distance spectrum of a linear time-invariant convolutional code $\CC$ may be derived from the weight distribution of a terminated code of length $N$ as $N \to \infty$.  

Without loss of generality, we may assume that $\CC$ is generated by a minimal encoder, which is necessarily \emph{noncatastrophic}:  \ie the unique state sequence associated with the all-zero code sequence is the all-zero state sequence.  Consequently, the lowest-weight words of a terminated code as $N \to \infty$ must be those that pass through the zero state almost all of the time.  These code sequences are as follows, for the various termination methods we have considered:
\begin{itemize}
\item If we terminate to the subcode $\CC_{[0,N)}$, then code sequences start and end in the zero state, and the lowest-weight sequences correspond to the lowest-weight sequences in the free distance spectrum.  If the minimum free distance is $\dfree$, then for $\dfree \le d < 2\dfree$ there will be approximately $N \times N_d$ sequences in the terminated code of weight $d$, where $N_d$ is the number of code sequences of weight $d$ in the free distance spectrum of $\CC$.  Thus, for $\dfree \le d < 2\dfree$, the weight distribution per unit time of $\CC$ is the limit of the weight distribution of $\CC_{[0,N)}$ normalized by (divided by) $N$ as $N \to \infty$.  For $d \ge 2\dfree$, there will be overcounting--- \eg two sequences of weight $\dfree$ may be counted as one of weight $2\dfree$--- but we will argue below that such overcounting should not affect estimates of code performance.
\item If we terminate to the projection $\CC_{|[0,N)}$, then code sequences can start and end in any state, and there will be low-weight sequences starting with a low-weight state transition $s \to 0$, remaining in state 0 for nearly $N$ time units, and then ending with a low-weight transition $0 \to s'$, where $s$ and $s'$ are not both 0.  Thus the minimum distance of  $\CC_{|[0,N)}$ will be less than $\dfree$ for all $N$.  However, the number of such low-weight sequences remains constant, so after normalization we will eventually see the same normalized weight distribution as for $\CC_{[0,N)}$.
\item If we terminate to the truncated code $\CC_{\lhd[0,N)}$, then by the same argument we will eventually see the same normalized weight distribution.  In this case the total weight of a code sequence starting in the zero state, remaining there for nearly $N$ time units, and then ending with a low-weight transition $0 \to s$, is only that of the low-weight transition  $0 \to s$.    However, again the number of such low-weight sequences remains constant, so after normalization we will eventually see the correct normalized weight distribution.
\item If we terminate to the tail-biting code $\CC_{||[0,N)}$, then by the same argument we will eventually see the same normalized weight distribution.  Note however that in this case the total weight of a code sequence starting with a low-weight transition $s \to 0$, remaining in the zero state for nearly $N$ time units, and then ending with a low-weight transition $0 \to s$, must be at least $\dfree$, since the ending sequence (corresponding to the state transition $0 \to s$) followed by the starting sequence (corresponding to $s \to 0$) must be a code sequence.  Thus the minimum distance of  $\CC_{||[0,N)}$ must equal $\dfree$ for large enough $N$.
\end{itemize} 

We conclude that as $N \to \infty$ the normalized weight distribution of any of these terminated codes approaches the free distance spectrum of $\CC$ for $\dfree \le d < 2\dfree$.  However, only the tail-biting termination has the same rate as $\CC$ and, for $N$ large enough, the same minimum distance $\dfree$.

Finally, we argue that the normalized weight distribution of any of these terminated codes $\CC_N$ must yield the same estimate of code performance over $N$ time units as the free distance spectrum of $\CC$, if these estimates are accurate.  The probability of error event $P(\E)$ of $\CC$ per unit time may be estimated using the free distance spectrum.  The probability of any error in $N$ time units is then estimated as $NP(\E)$.  If this is a good estimate (implying $N < 1/P(\E)$), then the probability of two or more error events in $N$ time units must be negligible.  But the probability of any error in decoding $\CC$ over $N$ time units is essentially the same as the probability of block decoding error in decoding $\CC_N$, which may be estimated by the weight distribution of $N$, which counts codewords that include two or more error events.  If the probability of two or more error events in $N$ time units is negligible, then an estimate based on the weight distribution of $\CC_N$ must approximately agree with an estimate based on the free distance spectrum of $\CC$.

\section{Weight adjacency matrices and weight generating functions}

We now show how Hamming weight generating functions for terminations of a linear time-invariant convolutional code $\CC$ may be derived from the Hamming weight adjacency matrix of a minimal linear time-invariant encoder for $\CC$.  This will allow us to state MacWilliams identities for terminated convolutional codes, and to estimate code performance.

Given a linear time-invariant encoder for a convolutional code $\CC$ with state space $\SSS$ and symbol alphabet $\A$, the Hamming weight adjacency matrix (HWAM) is the matrix $\Lambda(x)$ indexed by $\SSS \times \SSS$ whose elements are 
$$\Lambda_{ss'}(x) = \sum_{a \in \T(s, s')} x^{w(a)},$$ where $x$ is an indeterminate, $\T(s, s')$ is the subset of symbols $a \in \A$ such that the state/symbol transition (``branch") $(s, a, s') \in \SSS \times \A \times \SSS$ actually occurs in the encoder, and $w(a)$ is the Hamming weight of the symbol $a \in \A$.

\vspace{1ex}
\noindent
\textbf{Example 1} (cont.).  For the rate-1/2 binary convolutional code $\CC$ of Example 1, the HWAM of the encoder of Figure 1(a) is
$$
\Lambda(x) \quad  = \quad
\begin{array}{c|c|c|c|c|}
s/s' & 00 & 10 & 01 & 11 \\
\hline
00 & 1 & x^2 & 0 & 0 \\
\hline
10 & 0 & 0 & x & x \\
\hline
01 & x^2 & 1 & 0 & 0 \\
\hline
11 & 0 & 0  & x & x \\
\hline
\end{array}
$$
For the orthogonal code $\CC^\perp$, the HWAM of the encoder of Figure 1(b) is
$$
\hat{\Lambda}(x) \quad  = \quad
\begin{array}{c|c|c|c|c|}
s/s' & 00 & 10 & 01 & 11 \\
\hline
00 & 1 & 0 & x^2 & 0 \\
\hline
10 & x^2 & 0 & 1 & 0 \\
\hline
01 & 0 & x & 0 & x \\
\hline
11 & 0 & x  & 0 & x \\
\hline
\end{array}
$$
which in this case is the transpose of $\Lambda(x)$.
\qed \vspace{1ex}

It is shown in \cite{GLS08, GLS09, F09, F10} how the HWAM $\Lambda(x)$ of a minimal encoder for $\CC$ determines the HWAM $\hat{\Lambda}(X)$ of a minimal encoder for $\CC^\perp$ and \emph{vice versa} via a MacWilliams-type identity, but we will not need that result here.

\begin{figure*}[!t]
$$
\Lambda^4(x)  =
\left[
\begin{array}{cccc}
1 + 2x^5 + x^6 & x^2 + x^3 + x^4 + x^7 & x^3 + 2x^4 + x^5 & x^3 + 2x^4 + x^5  \\
x^3 + 2 x^4 + x^5 & x^2 + x^3 +  x^5 + x^6 & 2x^3 + x^4 + x^6 & 2x^3 + x^4 + x^6 \\
x^2 + x^3 + x^4 + x^7 & x^2 +  x^4 + 2x^5 & x^2 + x^3 + x^5 + x^6 & x^2 + x^3 + x^5 + x^6 \\
x^3 + 2 x^4 + x^5 & x^2 + x^3 +  x^5 + x^6 &2x^3 + x^4 + x^6 & 2x^3 + x^4 + x^6
\end{array}
\right].
$$
\large\caption{HWAM $\Lambda^4(x)$ of a section of $N = 4$ time units of Example 1 code $\CC$.}
\vspace{1ex}
\hrule
\end{figure*}

Now it is easy to see that if we take $N$ consecutive trellis sections of a minimal encoder for $\CC$ as a single section, then the HWAM of this length-$N$ trellis section is simply the $N$th power $\Lambda^N(x)$ of the basic HWAM $\Lambda(x)$.

\vspace{1ex}
\noindent
\textbf{Example 1} (cont.).  Given the HWAM $\Lambda(x)$ above for a minimal encoder of our example code $\CC$, the HWAM of a section consisting of $N = 2$ time units of this code is
$$
\Lambda^2(x) \quad  = \quad
\left[
\begin{array}{cccc}
1 & x^2 & x^3 & x^3 \\
x^3 & x & x^2 & x^2 \\
x^2 & x^4 & x & x \\
x^3 & x & x^2 & x^2 
\end{array}
\right].
$$
This shows that there is exactly one path from each state at time $k$ to each state at time $k+2$, and that the minimum Hamming weight of any of these paths (other than the zero path) is 1.

For a section consisting of $N = 4$ time units of this code, the HWAM $\Lambda^4(x)$ is given in Fig.\ 2.
This HWAM shows that there are four paths from each state  at time $k$ to each state at time $k+4$, and that the minimum nonzero Hamming weight of any of these paths is 2. \qed \vspace{1ex}

The weight generating functions of various terminated codes of $\CC$ can now be read from these weight adjacency matrices.  Since the subcode $\CC_{[0,N)}$ is the set of all sequences in $\CC$ that pass through the zero states at times 0 and $N$, its weight generating function is simply the $(0,0)$ element of $\Lambda^N(x)$.  Similarly, since the projection $\CC_{|[0,N)}$ is the set of all sequences in $\CC$ that pass through any states at times 0 and $N$, its weight generating function is the sum of all elements of $\Lambda^N(x)$.

Since the truncated code $\CC_{\lhd[0,N)}$ is the set of all sequences in $\CC$ that pass through the zero state at times 0 and any state at time $N$, its weight generating function is the sum of all elements in the first row of $\Lambda^N(x)$.  Similarly, the weight generating function of $\CC_{\rhd[0,N)}$ is the sum of all elements in the first column of $\Lambda^N(x)$.

Finally, since the tail-biting termination $\CC_{||[0,N)}$ is the set of all sequences in $\CC$ that pass through the same states in $S_0$ and $S_N$, its weight generating function is the sum of all diagonal elements of $\Lambda^N(x)$;  \ie its trace $\Tr(\Lambda^N(x))$.  Since $\CC_{||[0,N)}$ and $(\CC^\perp)_{||[0,N)}$ are orthogonal block codes, there is a MacWilliams identity between their weight generating functions.  

\vspace{1ex}
\noindent
\textbf{Example 1} (cont.).  For the rate-1/2 binary convolutional code of Example 1, the Hamming weight generating function of the the tail-biting termination $\CC_{||[0,4)}$ of length 4 is the trace of $\Lambda^4(x)$, namely $1 + 2x^2 + 4x^3 + x^4 + 4x^5 + 4x^6$.  Since $\hat{\Lambda}^4(x)$ is the transpose of $\Lambda^4(x)$, the orthogonal tail-biting terminated code $(\CC^\perp)_{||[0,4)}$ is an equivalent code with the same Hamming weight generating function.  It is easy to check that this Hamming weight generating function is indeed invariant under the MacWilliams transform. \qed \vspace{1ex}

Using tail-biting terminated codes, and normalizing the weight distribution by dividing by $N$, we have that the generating function of the normalized Hamming weight distribution of $\CC$ is
$$
g_{\CC}(x) = \lim_{N \to \infty} \frac{1}{N} \Tr(\Lambda^N(x)).
$$
Moreover, there is a MacWilliams identity between $g_{\CC}(x)$ and $g_{\CC^\perp}(x)$.  The performance of $\CC$ is determined by $g_{\CC}(x)$, and that of $\CC^\perp$ by $g_{\CC^\perp}(x)$.
(Similar observations are made in \cite{BHJK}, using truncated codes.)

\vspace{1ex}
\noindent
\textbf{Example 1} (cont.).  For a section consisting of $N = 16$ time units of the rate-1/2 binary convolutional code $\CC$ of Example 1, the HWAM $\Lambda^{16}(x)$ (modulo $x^8$) is given in Fig. 3 at the top of the next page.  Notice that 
$$\Tr(\Lambda^{16}(x)) = 1 + 16x^5 + 32x^6 + 64x^7 + \cdots,$$
 so that normalizing the distribution by dividing by $N = 16$ already gives the precise free distance spectrum of $\CC$ for $d < 8$, namely $x^5 + 2x^6 + 4x^7 + \cdots$. Thus the convergence to the limiting generating function $g_{\CC}(x)$ is rapid and exact.  This property of tail-biting terminations is not shared by other kinds of terminations. \qed \vspace{1ex}

\begin{figure*}[!t]\small
$$
\begin{array}{cccc}
1 + 14x^5 + 25x^6 + 44x^7 & x^2 + x^3 + 2x^4 +4x^5 +8x^6 + 29x^7 & x^3 + 2 x^4 + 4x^5 + 8x^6 + 16x^7 & x^3 + 2 x^4 + 4x^5 + 8x^6 + 16x^7  \\
x^3 + 2 x^4 + 4x^5 + 8x^6 + 16x^7 & x^5 + 3x^6 + 8x^7 & x^6 + 4x^7 & x^6 + 4x^7 \\
x^2 + x^3 + 2x^4 +4x^5 +8x^6 + 29x^7 & x^4 + 2x^5 +5x^6 +12x^7 & x^5 + 3x^6 + 8x^7 & x^5 + 3x^6 + 8x^7 \\
x^3 + 2 x^4 + 4x^5 + 8x^6 + 16x^7 & x^5 + 3x^6 + 8x^7 & x^6 + 4x^7 & x^6 + 4x^7
\end{array}
$$
\large\caption{HWAM $\Lambda^{16}(x)$ (modulo $x^8$) of a section of $N = 16$ time units of Example 1 code $\CC$.}
\vspace{1ex}
\hrule
\end{figure*}

It appears that the behavior of $g_{\CC}(x)$ might be analyzed by using an extension of Perron-Frobenius theory to generating function matrices, as in  \cite{FKMT}; however, we have not attempted such an analysis.


\vspace{1ex}
\noindent
\textbf{Example 2} (\cf \cite{SM77, BHJK}).  The two codes proposed by Shearer and McEliece \cite{SM77} for their counterexample make an excellent example.  The first code is a rate-1/3 binary linear time-invariant convolutional code $\CC_1$ generated by the degree-1 generators $(1, 1 + D, D)$, \ie $\CC_1$ is generated by a minimal encoder with impulse response $(110, 011, 000, \ldots)$, whose trellis section is shown in Figure 4(a).  The HWAM of this encoder is
$$
\Lambda_1(x) \quad  = \quad
\left[
\begin{array}{cc}
1 & x^2 \\
x^2 & x^2 \\
\end{array}
\right].
$$

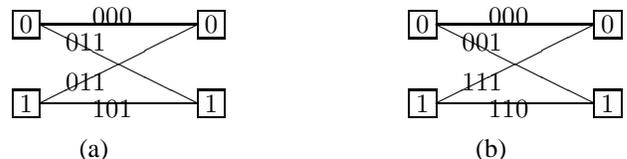
\begin{figure}[h]
\setlength{\unitlength}{5pt}
\centering
\begin{picture}(42,8)(2, -1)
\put(0,0){\framebox(2,2){$1$}}
\put(0,6){\framebox(2,2){$0$}}
\put(14,0){\framebox(2,2){$1$}}
\put(14,6){\framebox(2,2){$0$}}
\put(2,1){\line(1,0){12}}
\put(2,1){\line(2,1){12}}
\put(2,7){\line(2,-1){12}}
\put(2,7){\line(1,0){12}}
\put(6,0){$101$}
\put(4,5){$011$}
\put(6,7){$000$}
\put(4,2){$011$}
\put(5,-3){(a)}

\put(30,0){\framebox(2,2){$1$}}
\put(30,6){\framebox(2,2){$0$}}
\put(44,0){\framebox(2,2){$1$}}
\put(44,6){\framebox(2,2){$0$}}
\put(32,1){\line(1,0){12}}
\put(32,1){\line(2,1){12}}
\put(32,7){\line(2,-1){12}}
\put(32,7){\line(1,0){12}}
\put(36,0){$110$}
\put(34,5){$001$}
\put(36,7){$000$}
\put(34,2){$111$}
\put(35,-3){(b)}
\end{picture}

\caption{Trellis sections of (a) rate-1/3 2-state binary convolutional code $\CC_1$; (b) similar code $\CC_2$.}
\label{Fig3}
\end{figure}
The second code is a rate-1/3 binary linear time-invariant convolutional code $\CC_2$ generated by the degree-1 generators $(D, D, 1 + D)$, \ie $\CC_2$ is generated by a minimal encoder with impulse response $(001, 111, 000, \ldots)$, whose trellis section is shown in Figure 4(b).  The HWAM of this encoder is
$$
\Lambda_2(x) \quad  = \quad
\left[
\begin{array}{cc}
1 & x \\
x^3 & x^2 \\
\end{array}
\right].
$$

Since the weights of the $0 \to 0$ and $1 \to 1$ transitions are the same for $\CC_1$ and $\CC_2$, and since the sums of the weights of the $0 \to 1$ and $1 \to 0$ transitions are the same, it is evident that the weight distributions of the subcodes $(\CC_1)_{[0,N)}$ and $(\CC_2)_{[0,N)}$ are the same for all $N$, and that the free distance spectra of $\CC_1$ and $\CC_2$ are also the same.  For the same reason, the weight distributions of the tail-biting terminated codes $(\CC_1)_{||[0,N)}$ and $(\CC_2)_{||[0,N)}$ are the same for all $N$. 

However, the weight distributions of the projections $(\CC_1)_{|[0,N)}$ and $(\CC_2)_{|[0,N)}$ are not the same even for $N = 1$. It follows that the weight distributions of the subcodes $(\CC_1^\perp)_{[0,N)}$ and $(\CC_2^\perp)_{[0,N)}$ of their orthogonal codes $\CC_1^\perp$ and $\CC_2^\perp$ are not the same, and therefore that their free distance spectra are not the same;  this was the point of Shearer and McEliece \cite{SM77}.

On the other hand, since the weight distributions of the tail-biting terminated codes $(\CC_1)_{||[0,N)}$ and $(\CC_2)_{||[0,N)}$ are the same for all $N$, it follows that  \emph{the weight distributions of the tail-biting terminated codes $(\CC_1^\perp)_{||[0,N)}$ and $(\CC_2^\perp)_{||[0,N)}$ are the same for all $N$.}

Since the performance of $\CC_1^\perp$ and $\CC_2^\perp$ can be analyzed from these weight distributions, it follows that the performance of $\CC_1^\perp$ and $\CC_2^\perp$ is effectively the same, despite the difference in their free distance spectra.\footnote{Another way of reaching the same conclusion is to observe that $\CC_1$ and $\CC_2$ are equivalent under a simple time-invariant, finite-memory permutation.  Therefore $\CC_1^\perp$ and $\CC_2^\perp$ are actually equivalent under the same permutation, and thus must have precisely the same performance.} \qed \vspace{1ex}

\section{Conclusion}

In summary, similarly to \cite{BHJK}, but using tail-biting terminated codes, we have shown that there is a MacWilliams identity between the generating functions of the weight distributions per unit time of a linear convolutional code $\CC$ and its orthogonal code $\CC^\perp$, and that this distribution is as useful as the free distance spectrum for estimating code performance.  These results effectively resolve the puzzle posed by Shearer and McEliece \cite{SM77}.

\section*{Acknowledgment}  
For an advance copy of \cite{BHJK}, I am grateful to  R. Johannesson.

\end{document}